\newcommand*{\ARXIV}{}
\ifdefined\ARXIV
    \documentclass[twocol]{ametsocV6.1}
\else
    \documentclass{ametsocV6.1}
\fi

\usepackage[utf8]{inputenc}
\usepackage{makecell}

\title{Seamless lightning nowcasting with recurrent-convolutional deep learning}
\authors{Jussi Leinonen,\aff{a}\correspondingauthor{Jussi Leinonen, jussi.leinonen@meteoswiss.ch} 
Ulrich Hamann\aff{a} and
Urs Germann\aff{a} 
}
\affiliation{
\aff{a}{Federal Office of Meteorology and Climatology MeteoSwiss, Locarno-Monti, Switzerland}
}

\abstract{A deep learning model is presented to nowcast the occurrence of lightning at a five-minute time resolution $60$ minutes into the future. The model is based on a recurrent-convolutional architecture that allows it to recognize and predict the spatiotemporal development of convection, including the motion, growth and decay of thunderstorm cells. The predictions are performed on a stationary grid, without the use of storm object detection and tracking. The input data, collected from an area in and surrounding Switzerland, comprise ground-based radar data, visible/infrared satellite data and derived cloud products, lightning detection, numerical weather prediction and digital elevation model data. We analyze different alternative loss functions, class weighting strategies and model features, providing guidelines for future studies to select loss functions optimally and to properly calibrate the probabilistic predictions of their model. Based on these analyses, we use focal loss in this study, but conclude that it only provides a small benefit over cross entropy, which is a viable option if recalibration of the model is not practical. The model achieves a pixel-wise critical success index (CSI) of $0.45$ to predict lightning occurrence within $8\ \mathrm{km}$ over the $60$-min nowcast period, ranging from a CSI of $0.75$ at a $5$-min lead time to a CSI of $0.32$ at a 60-min lead time.}

\begin{document}

\maketitle

%
\statement
We have developed a method based on artificial intelligence to forecast the occurrence of lightning at five-minute intervals within the next hour from the forecast time. The method utilizes a neural network that learns to predict lightning from a set of training images containing lightning detection data, weather radar observations, satellite imagery, weather forecasts, and elevation data. We find that the network is able to predict the motion, growth and decay of lightning-producing thunderstorms, and that when properly tuned, it can accurately determine the probability of lightning occurring. This is expected to permit more informed decisions to be made about short-term lightning risks in fields such as civil protection, electric grid management and aviation.

\section{Introduction}

Lightning poses significant hazards to society, whether directly through lightning strikes on humans, or indirectly by, for instance, igniting fires, damaging electrical infrastructure or disrupting aviation. This causes significant harm to human health and life and considerable monetary costs \citep[e.g.][]{Holle2014Lightning,Holle2016Lightning}, and climate change is expected to continue to make these societal impacts worse \citep{Price1994CO2Lightning,Koshak2015LightningImpacts}. Conversely, lightning avoidance also incurs significant costs in the form of cancellations, delays and service outages. It is therefore important that these precautions be taken when required, but that unnecessary interventions be avoided when possible. The accurate prediction of lightning is required for making informed decisions regarding the proper response to its impending occurrence.

Lightning usually originates in convective storms, which develop rapidly and occur in limited areas. This makes it difficult to predict their exact location using numerical weather prediction (NWP). For short lead times, it is often preferable to use \emph{nowcasting}, the statistical prediction of the development of weather patterns using the most recent available observations. On the other hand, statistical nowcasting models, being ignorant of the physics of the atmosphere, begin to lose their predictive power at longer forecast timescales. Merging the two approaches by using the statistical nowcast for near-term predictions, the NWP forecast at longer lead times, and a combination of the two in-between, is known as \emph{seamless} nowcasting \citep{Kober2012Seamless,Wastl2018Seamless,Nerini2019Reduced,Sideris2020NowPrecip}.

As with many complex statistical data problems, lightning nowcasting has recently been the subject of research applying machine-learning (ML) models to this problem. The ML approaches can be broadly divided into two categories: \emph{object-based} nowcasting, which uses conventional methods to detect storm objects and their motion and then applies machine learning to predicting their development, and \emph{grid-based} nowcasting, which operates directly on gridded data and produces gridded outputs. Grid-based nowcasting avoids the complexity of the storm detection and tracking algorithms (including a the somewhat arbitrary cell definition and the problems with merging and splitting cells) at the cost of requiring more advanced machine learning methods, especially if it is also desired to predict the motion of thunderstorm cells. In the object-based category, recently published research includes the studies of \citet{Shrestha2021LightningNowcasting} and \citet{Leinonen2021DataSources}. Among studies on grid-based ML prediction of lightning, \citet{Lin2019LightningForecast}, \citet{Zhou2020LightningDL}, and \citet{Geng2021LightningForecast} used deep-learning techniques using convolutional neural networks (CNNs), while \citet{Blouin2016LightningPrediction} and \citet{LaFata2021LightningNowcasting} used methods based on decision trees. \citet{Mostajabi2019LightningNowcasting} considered the nowcasting of lightning at weather station locations using ML.

Our study falls into the category of grid-based nowcasting. We present a neural network with convolutional layers to model spatial features and recurrent layers to model temporal development. This network draws data from multiple sources including lightning detection, weather radar, satellite imagery, NWP and topographical information. The architecture of the network is specifically designed to seamlessly combine data from observational and NWP sources. The output of the network can be interpreted as the probability of lightning occurrence, and therefore is suitable for uncertainty quantification and for enabling each end user to select thresholds that best conform to their needs. Our work is similar to that of \citet{Geng2021LightningForecast}, but complements and improves upon their work with a larger set of input data and a neural network specifically designed to incorporate NWP data seamlessly and to retain high-resolution input features in the near-term predictions. We also use higher spatial and temporal resolution (1 km and 5 min compared to their 4 km and 60 min) and a shorter maximum lead time (1 h vs. 6 h), emphasizing very-short-term warnings of lightning hazards. Furthermore, we present a more extensive analysis of various model features, especially training loss functions and calibration.

In this article, we first describe the input datasets (Sect.~\ref{sect:data}) and the neural network model used to make the predictions (Sect.~\ref{sect:models}). We then compare the different network features, evaluate the best models and discuss examples of predicted cases of lightning (Sect.~\ref{sect:results}). Finally, we summarize the results and present final conclusions and directions for future work (Sect.~\ref{sect:conclusions}).

\section{Data} \label{sect:data}

\subsection{Study area}

\begin{figure}
    \ifdefined\ARXIV
        \centerline{\includegraphics[width=\linewidth]{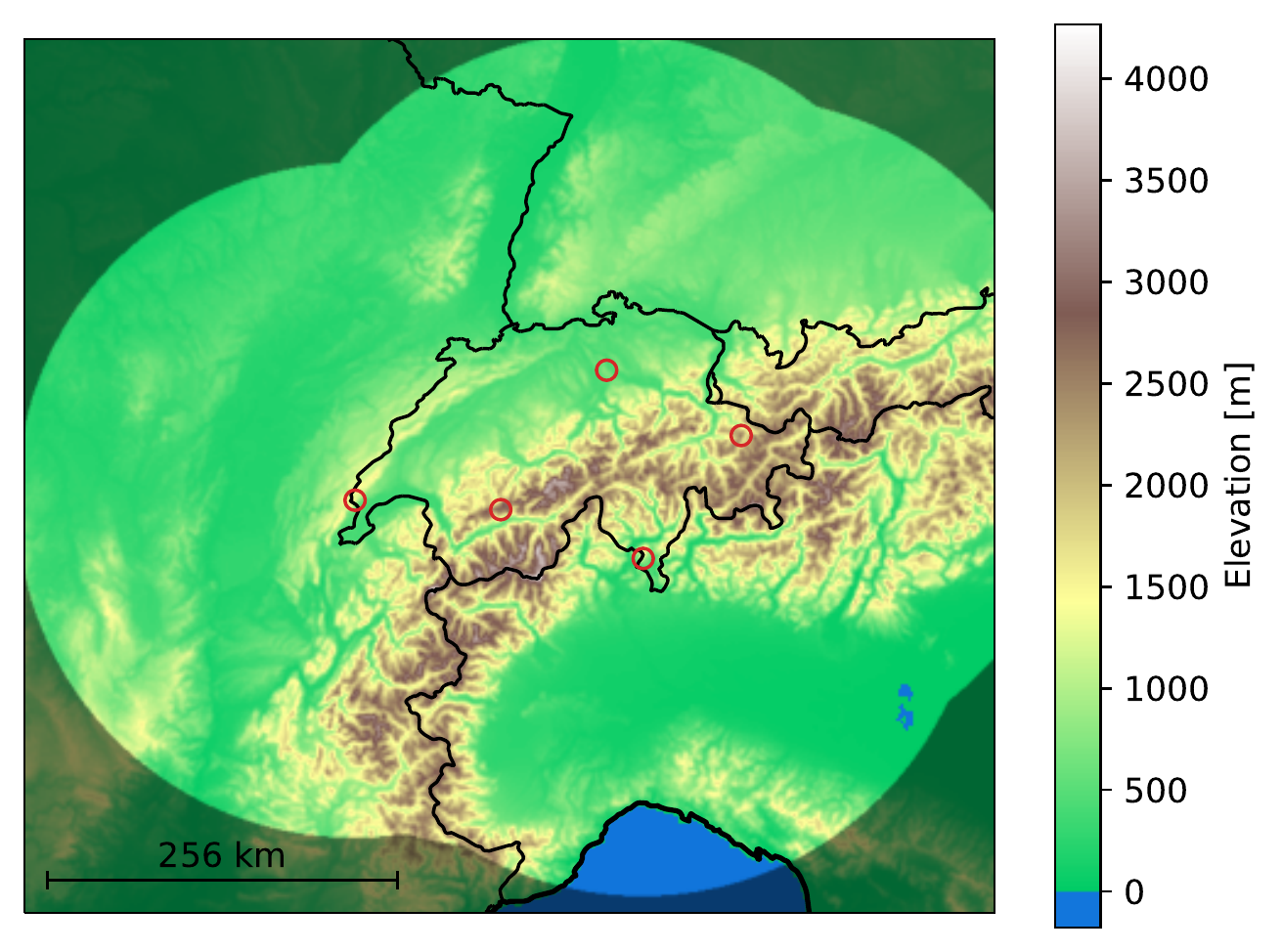}}
    \else
        \centerline{\includegraphics[width=0.8\textwidth]{study-area.pdf}}
    \fi
    \caption{The study area with the terrain elevation shown in color and the international borders as black lines. The locations of the weather radars are shown as red circles and the shaded area depicts the area outside the range of the radars and hence excluded from the study. The scale bar indicates a distance of 256 km, the size of the subdomains used for training.}
    \label{fig:study_area}
\end{figure}

We carried out our study in an area roughly defined by the coverage of the Swiss operational radar network, shown in Fig.~\ref{fig:study_area}. This area contains all of Switzerland as well as a considerable distance in each direction beyond its borders. We chose the area because it provides plentiful data for thunderstorm research, is covered comprehensively by radars with high spatial and temporal resolution, and has a high population density and sensitive infrastructure that make the prediction of severe weather particularly important. Another objective aided by the selection of this area was that of easing the process of adapting the research results into operational applications at a later stage. 

The study area is characterized by highly variable terrain. Terrain types range between the flat plains of the Po, Rhine and Saône valleys, the moderate-elevation regions of the Black Forest, the Vosges, the Jura, the French Massif Central and the Ligurian Apennines, and the high main chain of the Alps where the surface elevation frequently reaches above $3000\,\mathrm{m}$ above sea level. A small part of the area in the south is covered by the Mediterranean Sea. The climatological occurrence of lightning in the area, particularly on the southern flank of the Alps, is among the highest in Europe \citep{Taszarek2019ThunderstormClimatology}, making it highly suitable for the purposes of this research. Thunderstorms are also frequent on the northern flank of the Alps. Comparing the occurrence, characteristics and driving forces of thunderstorms on the northern and southern sides reveals similarities but also differences, as has been shown for hail \citep{Nisi2018Hail,Barras2021Hail}.

We process our data on a regular grid that covers the study area with $1\ \mathrm{km}$ resolution. The grid is defined using the EPSG:21781 projection, covering the range $[{255}\ \mathrm{km},{965}\ \mathrm{km}]$ in the projection coordinates in the east--west direction and $[{-160}\ \mathrm{km},{480}\ \mathrm{km}]$ in the north--south direction, resulting in a grid of $710 \times 640$ points. Some products such as the radar composite are natively produced on this grid; other products were projected into this grid before further processing. The regions out of range of the radars, shown shaded in Fig.~\ref{fig:study_area}, are excluded from the study.

\subsection{Data sources and preprocessing}

Below, we describe the data sources used in this study, how we expect them to contribute to the prediction of lightning, and the preprocessing applied to them. A complete list of the input variables can be found in Appendix Tables~\ref{table:preprocessing-1} and~\ref{table:preprocessing-2}.

\subsubsection{Lightning detection}

The current lightning activity is an excellent predictor for the occurrence of lightning in the near future. Our lightning data were collected with the European Cooperation for Lightning Detection (EUCLID) network of ground-based lightning detection antennas that determine the location of lightning strikes using triangulation and time-of-arrival differences. The operation of the network is described by \citet{Schulz2016EUCLID} and \citet{Poelman2016EUCLID}. The data were processed and provided to MeteoSwiss by Météorage. 

The raw data products consist of the time, location, estimated current and various other descriptors of individual lightning strikes. To transform these data into a format compatible with our network, we accumulated the individual strikes into maps of lightning density covering five-minute periods. We also created similar maps of current-weighted density. These maps were normalized to bring the mean activity close to $1$. Furthermore, we used the lightning data to create our targets, where a pixel is set to $1$ if a lightning strike occurred within $8\ \mathrm{km}$ of that pixel within the last $10\ \mathrm{min}$, and otherwise to $0$. This definition is used in safety procedures at airports for takeoff and landing operations based on the regulations of the \citet{EUAirTraffic2017} and the \citet{ICAOAirTraffic2018}. Such a binary definition is not concerned with the lightning flash rate but the occurrence of lightning regardless of the rate, and therefore emphasizes cases with low flash rate. Such situations are particularly hazardous because lightning may surprise people who have not yet sought shelter \citep{Holle1993Lightning}. This definition is used here to demonstrate the algorithm for one realistic use case, but the definition can be easily modified to accommodate the needs of other users. For example, to emphasize low flash rate situations further, the radius used to define lightning occurrence could be increased from $8\ \mathrm{km}$ or the time window widened from $10\ \mathrm{min}$.

\subsubsection{Precipitation radar}

The precipitation radar data originate from the Swiss operational weather radar network operated by MeteoSwiss \citep{Germann2016SwissRadar,Germann2022RadarOrography}. The network consists of five C-band dual-polarization Doppler radars, which cover the entire area of Switzerland along with regions of the neighboring countries. The radars indirectly measure the intensity of precipitation at the surface. As the radars scan at multiple elevation angles and the scans of different radars overlap, the radars also provide information on the three-dimensional structure of the radar reflectivity. The relatively short distances between the radars compared to their range, and the strategic placement of certain radars at high elevations, mitigate the observational gaps caused by the terrain blocking the radar beams. The operational processing chain merges the measurements from the different radars into one composite on the study grid.

Of the radar data, we used the vertical-profile-corrected estimate of the precipitation at the surface (RZC), the maximum column reflectivity (CZC), echo top heights at radar reflectivity thresholds of 20 dBZ and 45 dBZ (EZC-20 and EZC-45, respectively), the vertically integrated liquid water content (LZC) and the height of the maximum radar echo (HZC). These were chosen from among the descriptors available in the radar archive because the radar reflectivity and its vertical profile provide a direct observation of the hydrometeors relevant for the initiation of lightning. In particular non-sticky collisions between graupel (or larger ice crystals) coated with a quasi-liquid layer of super-cooled water and smaller upward moving ice crystals in the convective updraft region cause a separation of electric charges \citep{Takahashi1978Riming}. The chosen radar observations have been identified by many sources as an indicator of thunderstorms and lightning \citep[e.g.][]{Marshall1978RadarLightning,Gremillion1999LightningRadar,Hering2004Nowcasting,Hering2006Operational,Houze2014CloudDynamics}. For processing in the neural network, we transformed RZC and LZC to a logarithmic scale, motivated by the globally lognormal distribution of rain intensity \citep{Kedem1987Lognormality}. We shifted and scaled RZC, CZC and LZC to distributions that are close to the standard normal distribution. EZC and HZC, which have natural minima at $0$, were scaled to a mean of approximately $1$. The details of the transformations can be found in Appendix Table~\ref{table:preprocessing-1}.

\subsubsection{Satellite imagery: SEVIRI}

The satellite data were obtained from the Rapid Scan service from the Spinning Enhanced Visible and InfraRed Imager \citep[SEVIRI;][]{Schmid2000SEVIRI}, which are found on board each of the MeteoSat Second Generation (MSG) satellites. The products we used originate from the MSG-3 satellite, which is on geostationary orbit over $0^{\circ}$ longitude. Every five minutes, the Rapid Scan service scans one third of the disk visible from the satellite, centered on Central and Western Europe. The SEVIRI instrument provides data at 12 narrow-wavelength bands ranging from the visible to the infrared. The details of the bands can be found in Appendix Table~\ref{table:preprocessing-1}, where the three numbers in each band name indicate the band wavelength: For instance, IR-087 corresponds to a band at $8.7$~\textmu{m}. Furthermore, SEVIRI produces a broadband high-resolution visible wavelength data product (HRV). Reflectance, brightness temperatures, as well as differences and temporal derivatives of the MSG bands are associated with convective cloud properties, as discussed by \citet{Mecikalski2010MSGConvectionPartI,Mecikalski2010MSGConvectionPartII}. Higher-level cloud data products are derived from the SEVIRI data by the Nowcasting Satellite Application Facility (NWCSAF). These include information about the type, height and microphysical properties of clouds. We included them as we expected them to convey further information about cloud-top phenomena associated with severe weather, such as overshooting tops \citep{Bedka2010Objective,Bedka2011Overshooting} and above anvil cirrus plumes \citep{Bedka2018Above}.

We input transformed versions of each of the SEVIRI bands into our network. The thermal bands are available at all times of day and are expressed as brightness temperatures, which were transformed to mean $\mu \approx 0$ and standard deviation $\sigma \approx 1$ using the same scaling for all bands except for IR-039, which is sensitive to both solar and infrared radiation, and hence was given its own scaling parameters.

The solar bands include HRV, the two visible-wavelength bands, as well as IR-016, the infrared band nearest to the visible range. These are not available at night, and are set to zero during these times. To cancel out diurnal variation, we divided the solar band radiances by $\cos \theta_\mathrm{z}$, where $\theta_\mathrm{z}$ is the solar zenith angle. After this, we applied thresholds below which the radiance was set to zero in order to mask out signals originating from the surface. While this may also hide some thin clouds, we do not expect this to be a major issue for the present application as we concentrate on phenomena that are associated with very thick clouds. Finally, we transformed the solar bands to $\mu \approx 1$.

Out of the NWCSAF products, we use the cloud phase, cloud top temperature, cloud top height and cloud optical thickness, the last of which is not available at night while the others are available at all times of day. Combinations of these have been used to identify deep convective (i.e.\ tall and optically thick) clouds in previous studies \citep[e.g.][]{Oreopoulos2014CloudRegimes}. The cloud phase is a categorical variable indicating either no cloud, liquid cloud, ice cloud or mixed-phase cloud. This was transformed into a one-hot feature. The cloud top temperature and height were scaled close to $(\mu,\sigma)\approx(0,1)$. Meanwhile, we took the logarithm of the cloud optical thickness following \citet{Leinonen2016CloudRetrieval} and normalized this to near $(\mu,\sigma)\approx(0,1)$.

The sub-satellite resolution of all satellite channels and products except HRV is ${3}\ \mathrm{km}$, corresponding to roughly ${3}\ \mathrm{km} \times {5}\ \mathrm{km}$ at the latitude range of the study area. These products were resampled to a grid with a resolution of ${4}\ \mathrm{km}$, shared with the NWP products. The HRV has a sub-satellite resolution of ${1}\ \mathrm{km}$, and was resampled to the ${1}\ \mathrm{km}$ resolution grid also used for the radar and lightning observations. The resampling was performed using projection to the ${1}\ \mathrm{km}$ grid with PyTroll \citep{Raspaud2018Pytroll}, followed by taking the average of each ${4}\ \mathrm{km} \times {4}\ \mathrm{km}$ square to reduce the resolution of the channels other than HRV.

We recognized that the lack of availability of the solar bands and the cloud optical thickness at night was a potential issue that might confuse the machine-learning algorithm to consider the scene cloudless. To provide information about the time of day, we also provided $\cos \theta_\mathrm{z}$ as an input. In principle, this should provide enough information to the network to enable it to disregard the solar bands when $\theta_\mathrm{z} \leq 0$.

\subsubsection{Numerical weather prediction}

The NWP products were derived from the archived operational forecast runs of the Consortium for Small Scale Modelling (COSMO) model \citep{Baldauf2011COSMO}, which MeteoSwiss uses for operational NWP. Analysis products would also have been available in the archive, but were not used as they would not be available in real-time operations.

Using the results of \citet{Leinonen2021DataSources} as a guideline, we selected various features that pertain to the occurrence of deep convection from among the COSMO model outputs. These were the convective available potential energy with respect to the most unstable level (CAPE-MU), the convective inhibition (CIN), the height of the ${0}\ \mathrm{^{\circ} C}$ isotherm (HZEROCL), the lifting condensation level (LCL), the moisture convergence (MCONV), the vertical velocity of air in pressure coordinates (OMEGA), the surface lifted index (SLI), the soil type, and the temperatures at the surface and at 2 m height (T-SO and T-2M, respectively). The COSMO model produces more variables that would be potentially useful for our prediction tasks. Unfortunately, some operational forecast output archives had only been retained for a limited time due to the vast data volume created, and it was not possible to recover all potentially interesting variables at the time the data collection was performed.

As with the other data sources, the features that have a natural zero point, such as those expressing height from the surface, were scaled to $\mu \approx 1$. Those with an essentially open data range, for example the temperatures, were shifted and scaled to approximately $(\mu,\sigma)\approx(0,1)$. The soil type is a categorical variable, and accordingly we transformed it into a one-hot feature.

The native resolution of the COSMO-1 version used operationally in Switzerland is $1.1\ \mathrm{km}$. We expected that the operational forecast would provide the largest benefits at longer lead times, when the spatial uncertainty of forecasted events is quite high. With this consideration, and since it was desirable to constrain the amount of data passed to the model, we downsampled the data to $4\ \mathrm{km}$ resolution by averaging $4 \times 4$ pixel squares after projecting the data to the study grid. The forecast products were available at a time resolution of 1 h; linear interpolation was used to produce frames at a 5 min resolution. Reflecting the expected operational use pattern, the latest NWP forecast with lead time of at least 1 hour was selected for each time step. Thus, the lead time of the NWP forecast used in the ML model ranges from 1 to 4 hours; information about the lead time was not passed to the ML model.

\subsubsection{Digital elevation model}

The elevation in parts of the study area is considerable, indeed the highest in Europe outside the Caucasus, and orography is widely known to physically influence convective processes \citep{Kirshbaum2018OrographicConvection} and to be statistically linked to lightning occurrence \citep[e.g.][]{Dissing2003Spatial}. Therefore, we expected that it would be important to include information about the elevation in the analysis despite the results of \citet{Leinonen2021DataSources} that suggested that the DEM does not contribute significantly to the prediction skill. We used a set of DEM data derived from the Advanced Spaceborne Thermal Emission and Reflection Radiometer (ASTER) global DEM \citep{Abrams2020ASTERV3}. Of the DEM features, we passed the elevation and the east--west and north--south direction derivatives to the model. The elevation was scaled to a mean of $1$ and the derivatives to $(\mu,\sigma)\approx(0,1)$.

\subsection{Selection and processing} \label{sect:preprocessing}

As we wanted to focus on predicting lightning, we downselected data from our study area and time period into the dataset such that only those spatiotemporal regions where convective activity was likely occurring nearby were included. We identified these regions based on the radar-derived rainfall rate. At each time step in the study period, we located regions in the study area where the rainfall rate exceeded ${10}\ \mathrm{mm\,h^{-1}}$ in a contiguous area of more than $10$ pixels. For each such area, we added to the dataset the spatiotemporal box containing the $256 \times 256$ pixels surrounding the area at every time step $\pm {2}\ \mathrm{h}$ from the time of occurrence. Although lightning can occur at rain rates lower than ${10}\ \mathrm{mm\,h^{-1}}$ or even with no rain \citep[e.g.][]{Hodanish2004UpdraftLightning,Schultz2021MesoscaleLightning}, the ${10}\ \mathrm{mm\,h^{-1}}$ threshold concerns the \emph{maximum} in the $256\ \mathrm{km} \times 256\ \mathrm{km}$ box; therefore, the training data should also contain many cases of lightning with lower rain rates and consequently the threshold should not overly restrict the training dataset.

After the downselection, the volume of the data was still prohibitively large for the typical approach of storing the training dataset in a single static tensor. However, the sequences in different training samples overlap spatially and temporally. To avoid duplication of data storage, we divided the study area into tiles of $32 \times 32$ pixels such that each tile is stored at most once on a given time step. Only tiles that are within the regions of interest defined by the rainfall rate threshold were stored. The training samples are generated on demand from these tiles during network training and evaluation. The elimination of data duplication allows us to greatly reduce the memory footprint or, conversely, use a larger amount of data within the limits of the available memory.

We divided the dataset into a training set used to train the model, a validation set used for performance evaluation at training time and for hyperparameter tuning, and a test set used for final evaluation. Following the strategy of \citet{Leinonen2021DataSources}, entire days were set aside for the training set and the testing set in order to minimize overlap between the three subsets of data. First, days were randomly selected until at least $10\%$ of the available time steps had been assigned into the validation set. Then, the same process was repeated with the remaining data to assign another at least $10\%$ into the testing set. The remaining data were used as the training set.

The final dataset includes a total of $30\,641$ different possible starting times for the training sequences, of which $24\,113$ are in the training set, $3210$ in the validation set and $3318$ in the testing set. In total, $1\,021\,447$ different samples, i.e.\ around $30$ samples per starting time, can be generated (not including the further diversity added by data augmentation). However, there is considerable overlap between these both spatially and temporally. The effective number of unique training samples in the dataset can be estimated from the total volume of the data: The total number of data points is approximately $6.7 \times 10^9$, which corresponds to $5680$ image sequences of $18 \times 256 \times 256$ size. Approximately $7.7 \times 10^7$ data points fulfill the condition that lightning occurred within 8 km in the previous 10 min; this is roughly $1.1\%$ of the total, indicating a severely unbalanced dataset. The generation of training samples will be discussed in more detail in Sect.~\ref{sect:models}\ref{sect:training}.

\section{Models} \label{sect:models}

\subsection{Neural network} \label{sect:network}

The model we propose for lightning nowcasting is a neural network that uses convolutional layers to model spatial relationships and recurrent connections to model the temporal evolution of the state of the atmosphere. The network follows an encoder-forecaster architecture, where the encoder produces an analysis of the state of the atmosphere as a deep representation, while the forecaster decodes this representation into a prediction of the future evolution of the target variable. Shortcut connections similar to U-Net networks \citep{Ronneberger2015UNet} are used to allow the encoder to be connected to the forecaster simultaneously at multiple different scales. The architecture was developed from that used by \citet{Leinonen2020Downscaling}; a variant developed in parallel work was used by \citet{Leinonen2021Weather4castBigData,Leinonen2021Weather4castStage1}. The design resembles those already used in nowcasting applications by \citet{Franch2020Nowcast}, \citet{Cuomo2021DLNowcasting} and \citet{Ravuri2021GenerativePrecipitation}.

\begin{figure*}
    \centerline{\includegraphics[width=\textwidth]{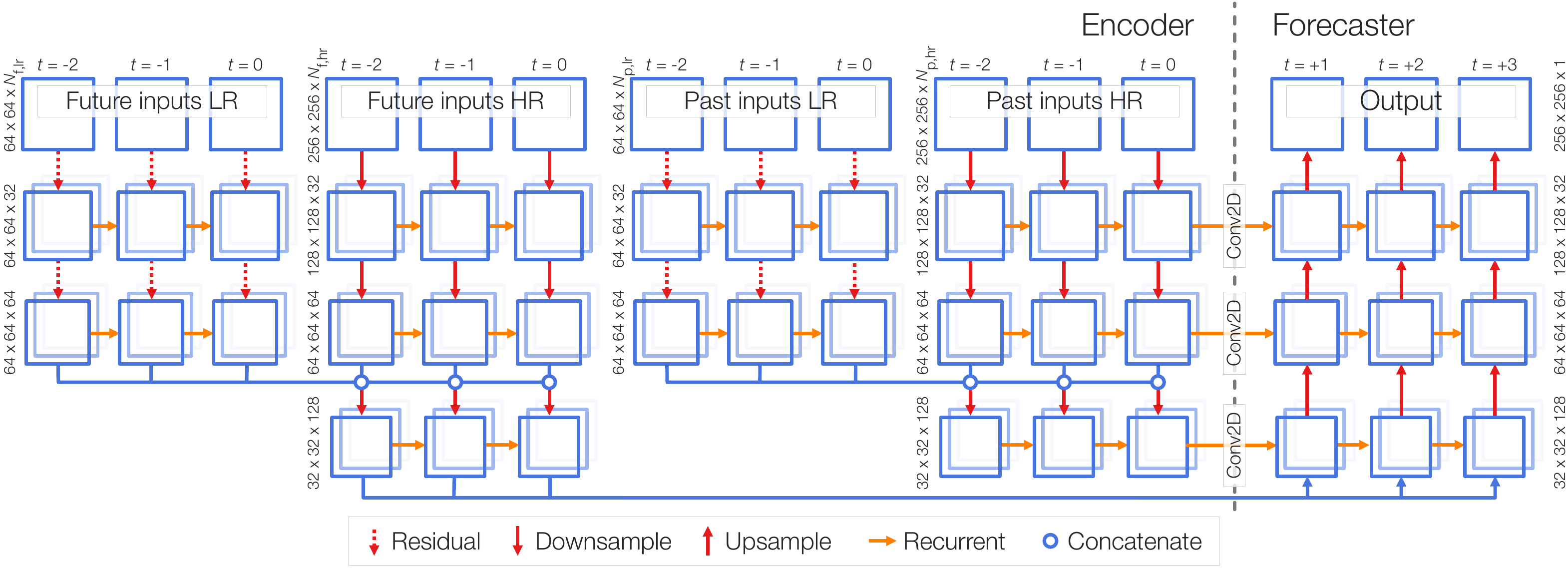}}
    \caption{An illustration of the network architecture. For clarity, only $3$ time steps are shown for both the past and the future; the actual network processes $6$ past time steps and $12$ future time steps. $N$ is the number of predictors; $\mathrm{lr}$ indicates low resolution and $\mathrm{hr}$ high resolution, while $\mathrm{p}$ indicates the past timeframe and $\mathrm{f}$ the future timeframe (i.e. COSMO variables). In our case $N_{\mathrm{f},\mathrm{lr}}=9$, $N_{\mathrm{f},\mathrm{hr}}=10$, $N_{\mathrm{p},\mathrm{lr}}=20$ and $N_{\mathrm{p},\mathrm{hr}}=20$.}
    \label{fig:network}
\end{figure*}

A diagram of the network architecture is shown in Fig.~\ref{fig:network}. The downsampling connections use residual blocks derived from ResNet \citep{He2016ResNet}, with strided convolutions to downsample the resolution by a factor of $2$. The upsampling connections are similar, but they apply bilinear upsampling by a factor of $2$ before the residual block. The recurrent connections use a variant of the convolutional gated recurrent unit \citep{Ballas2016ConvGRU} where the convolutions have been replaced with residual blocks as in \citet{Leinonen2021Weather4castStage1}. The weights of the downsampling blocks in the encoder are shared between the time steps, as are those of the upsampling blocks in the forecaster. The recurrent connections at a given resolution use shared weights in the encoder and shared weights in the forecaster, but the weights are not shared between the encoder and the forecaster. The hidden states of the recurrent units in the encoder are initialized to zero, while those in the forecaster are initialized to the final state of the corresponding recurrent unit in the encoder, passed through a convolutional layer to allow the forecaster to have representations different from those of the encoder. The output of the forecaster is a time series of images with each pixel set to a value between $0$ and $1$, with larger values corresponding to higher confidence in lightning occurrence at that location and time step.

The encoder has multiple branches that correspond to the different input data time frames (past or future) and spatial resolutions. When a prediction is made, observational data are clearly only available for the past time frame, but NWP forecasts are also available for the future and can be exploited to produce a seamless nowcast. The output of the future branch of the encoder, which uses the NWP data, is used as the input to the deepest recurrent layer of the forecaster. The branches corresponding to different spatial resolutions are treated equally except that the downsampling operations are skipped in the lower-resolution branches until the resolution becomes equal to the next-highest resolution branch, at which point the two branches are merged using concatenation. The network can create past branches with the full resolution and additional resolutions $2$, $4$ and $8$ times lower, and future branches with the same set of resolutions. The network construction script analyzes the inputs and automatically creates only the branches that are necessary for the input data. In the case of our input data, we have two past branches with resolutions of ${1}\ \mathrm{km} \times {1}\ \mathrm{km}$ (radar, lightning, DEM and HRV data) and ${4}\ \mathrm{km} \times {4}\ \mathrm{km}$ (SEVIRI data other than HRV) and two future branches with the same resolutions (NWP data at ${4}\ \mathrm{km} \times {4}\ \mathrm{km}$ and static data such as the DEM at ${1}\ \mathrm{km} \times {1}\ \mathrm{km}$). An architecturally simpler alternative would have been to upsample all data to the highest resolution before processing. Processing the lower-resolution input data at a resolution closer to the original saves processing time, memory and CPU-to-GPU transfer time. In the future, the model can also easily be adapted to new predictor datasets. Due to its capacity to process datasets with different resolutions, the model will be well suited to process the observations from the upcoming Meteosat Third Generation, where solar observations are made with 1 km sub-satellite resolution and thermal channels with 2 km resolution.

The model in its baseline configuration did not use normalization or dropout. The experience from \citet{Leinonen2021Weather4castStage1} was that the model architecture is resilient to overfitting and these mitigation tools are not necessarily needed. Dropout can be optionally included and the consequences of this will be discussed in Sect.~\ref{sect:results}\ref{sect:selection}.

The network was implemented in Python using Tensorflow/Keras \citep{Chollet2015Keras}. NumPy \citep{Harris2020NumPy}, SciPy \citep{Virtanen2020Scipy} and the PyTroll libraries \citep{Raspaud2018Pytroll} were used for data processing, Numba \citep{Lam2015Numba} and Dask \citep{Rocklin2015Dask} for optimization and Matplotlib \citep{Hunter2007Matplotlib} for visualization.

\subsection{Training} \label{sect:training}

We trained the model to predict $N_\mathrm{f}=12$ time steps in the future from observational data from the $N_\mathrm{p}=6$ time steps in the past and NWP data for the $N_\mathrm{f}$ future time steps. At the 5 min time resolution, this corresponds to 30 min in the past and 60 min in the future. We briefly examined using a length of $12$ time steps also for the past time frame, but found no noticeable benefits over $6$ steps. 

At initialization, the data generator locates all available three-dimensional boxes of ($N_\mathrm{p}+N_\mathrm{f}) \times N_\mathrm{h} \times N_\mathrm{w}$ pixels (where $N_\mathrm{h}=N_\mathrm{w}=256$ are the height and width of the training images, respectively) in the 1 km resolution training data. These have been selected to be in or near regions containing likely convection as described in Sect.~\ref{sect:data}\ref{sect:preprocessing}. During each training epoch, the data generator iterates over all possible starting times of the training sequences in randomized order. Only one training sample is generated for each starting time at each epoch, selected randomly from among the possible choices. This reduces the overlap of training samples, with the objective of avoiding overfitting to the cases of most widespread convection. Additionally, random rotations at $90^{\circ}$ intervals and random mirroring are used to further increase the diversity of training samples and to incentivize the network to learn to be approximately invariant with respect to these transformations.

The training was performed on a computing cluster node with eight Nvidia V100 GPUs. With this hardware, one epoch required approximately ${18}\ \mathrm{min}$ of training time. The number of training epochs was not fixed, as different loss functions may require different amounts of training time. Instead, we followed an early stopping strategy where the learning rate is divided by $5$ if the loss in the validation set has not improved for three epochs, and the training is stopped if the validation loss has not improved for six epochs; after stopping, the model weights giving the best validation loss are saved. Unlike with the training set, the order of samples in the validation and testing sets was not shuffled, nor was random data augmentation applied, in order to prevent a spurious improvement in the validation and test losses due to a random selection of more favorable inputs. We found that the training typically stopped after $20$--$30$ epochs.

In contrast to the training time, the model is quite fast to evaluate: we were able to generate a single sample and produce a prediction for it in $1.2\ \mathrm{s}$ on a modern computer with 16 CPU cores. Thus, GPU hardware is not necessary to use the model operationally, and bottlenecks in producing warnings are likely to be in data acquisition rather than computation.

\subsection{Evaluation} \label{sect:evaluation}

Various skill scores can be computed to describe the predictive power of a binary forecast model; these have been summarized in the atmospheric science context by \citet{Hogan2012BinaryForecasts}. Scores differ in terms of whether they assign more weight to the correct prediction of occurring events or that of non-occurrence. Moreover, for highly unbalanced datasets such as that used in our study, some skill scores are less suitable than others \citep{Branco2016SurveyImbalanced}.

A probabilistic forecast can be turned into a deterministic yes/no forecast by selecting a decision threshold $T$, above which it is predicted that an event will occur, while otherwise a prediction of non-occurrence is issued. Such skill scores are computed from the confusion matrix (also called contingency table), which divides the predictions into four cases: true positives ($\mathrm{TP}$), false positives ($\mathrm{FP}$), false negatives ($\mathrm{FN}$) and true negatives ($\mathrm{TN}$). These can be logically defined as
\begin{eqnarray}
\mathrm{TP} &=& \frac{1}{N} \sum_{i=1}^N {\hat y}_i \land y_i \label{eq:tp}\\
\mathrm{FP} &=& \frac{1}{N} \sum_{i=1}^N {\hat y}_i \land \neg y_i \label{eq:fp}\\
\mathrm{FN} &=& \frac{1}{N} \sum_{i=1}^N \neg {\hat y}_i \land y_i \label{eq:fn}\\
\mathrm{TN} &=& \frac{1}{N} \sum_{i=1}^N \neg {\hat y}_i \land \neg y_i \label{eq:tn}
\end{eqnarray}
where $\neg$ represents the logical ``not'', and $\land$ the logical ``and'', $y_i$ are the observed events, ${\hat y}_i$ are the predictions and $N$ is the total number of data points. In the definitions above, we have normalized these such that $\mathrm{TP}+\mathrm{FP}+\mathrm{FN}+\mathrm{TN}=1$. It can be easily verified that the scores presented below are not affected by this normalization.

From the scores defined in Eqs.~\ref{eq:tp}--\ref{eq:tn}, one can derive various relatively straightforward metrics of success. Many of the metrics have varying names in different fields; below we define each metric only once and mention the alternative names.
\begin{description}
\item[Probability of detection (POD)] gives the fraction of occurrences that were predicted:
\begin{equation}
    \mathrm{POD} = \frac{\mathrm{TP}}{\mathrm{TP}+\mathrm{FN}} \label{eq:pod}
\end{equation}
POD is also known as \emph{recall} or \emph{true positive rate}.
\item[False alarm ratio (FAR)] gives the fraction of predicted occurrences where the event did not, in fact, occur:
\begin{equation}
    \mathrm{FAR} = \frac{\mathrm{FP}}{\mathrm{TP}+\mathrm{FP}}
\end{equation}
The complement $1-\mathrm{FAR}$ is called \emph{precision}.
\item[False positive rate (FPR)] gives the fraction of non-occurrences of the event that were incorrectly predicted as occurrences:
\begin{equation}
    \mathrm{FPR} = \frac{\mathrm{FP}}{\mathrm{FP}+\mathrm{TN}}
\end{equation}
This is also known as the false alarm \emph{rate}, but we avoid this name in order to reduce confusion with the FAR.
\end{description}
The scores above are tradeoffs with respect to each other: For example, a decision threshold of $T=0$ gives a POD of $1$ but a high FAR, while $T=1$ gives a FAR of $0$ but also a POD of $0$. For this reason, other scores have been devised which balance the different correct and incorrect predictions and generally attain their maximum at some threshold $0<T<1$. Of these, we use the following:
\begin{description}
\item[Critical success Index (CSI)] expresses the fraction of observations or predictions that were correct:
\begin{equation}
    \mathrm{CSI} = \frac{\mathrm{TP}}{\mathrm{TP}+\mathrm{FP}+\mathrm{FN}} . \label{eq:CSI}
\end{equation}
This score is also known in meteorology as the \emph{threat score}, and more widely as the \emph{Jaccard index} or the \emph{intersection-over-union score}. The latter term arises from the fact that the numerator in Eq.~\ref{eq:CSI} can be interpreted of the intersection of the sets of observed and predicted events, while the denominator can be interpreted as their union.
\item[Equitable threat score (ETS)] is a adjustment of the CSI that measures the skill of a prediction relative to a random forecast $R$:
\begin{eqnarray}
    R & = & \frac{(\mathrm{TP}+\mathrm{FN}) \cdot (\mathrm{TP}+\mathrm{FP})}{\mathrm{TP}+\mathrm{FP}+\mathrm{TN}+\mathrm{FN}} \\
    \mathrm{ETS} & = & \frac{\mathrm{TP}-R}{\mathrm{TP}+\mathrm{FP}+\mathrm{FN}-R}
\end{eqnarray}
\item[Heidke skill score (HSS)] is another measure of the improvement of the fraction of correct forecasts over that given by a random (skilless) forecast:
\begin{equation}
    \mathrm{HSS} = \frac{2 (\mathrm{TP} \cdot \mathrm{TN} - \mathrm{FN} \cdot \mathrm{FN})}{(\mathrm{TP}+\mathrm{FN}) \cdot (\mathrm{FN}+\mathrm{TN}) + (\mathrm{TP}+\mathrm{FP}) \cdot (\mathrm{FP}+\mathrm{FN})} . \label{eq:HSS}
\end{equation}
\item[Peirce skill score (PSS)] measures the separation of positive and negative occurrences, defined as
\begin{equation}
    \mathrm{PSS} = \frac{\mathrm{TP}}{\mathrm{TP}+\mathrm{FN}} - \frac{\mathrm{FP}}{\mathrm{FP}+\mathrm{TN}} . \label{eq:PSS}
\end{equation}
\end{description}

The above scores give the performance of a model for a given threshold $T$. A probabilistic forecast can also be evaluated using its overall performance over the entire range of possible thresholds $0 \leq T \leq 1$. Two prominent examples of this involve calculating an area under a curve (AUC) by integration. The \emph{receiver operating characteristic} (ROC) AUC considers the curve of the POD as a function of the FPR, both of which increase with decreasing $T$. The ROC AUC is often not very informative for severely unbalanced datasets like ours. For these, the \emph{precision--recall} (PR) AUC is recommended instead \citep{Davis2006PRROC}. This is computed from the curve of the precision as a function of the recall, that is, $1-\mathrm{FAR}$ as a function of the POD. The AUC characteristics can also be graphically inspected by plotting the curves.

\subsection{Training losses} \label{sect:loss}

Our prediction task falls into the general category of predicting the probability of a binary event occurring. As the probability is predicted for each pixel, the prediction also has much in common with image segmentation tasks, where the task of a model is to identify the pixels in an image that belong to a certain category. Loss functions for image segmentation have been systematically reviewed recently by \citet{Jadon2020SegmentationLoss} and \citet{Mehrtash2020Segmantation}. One important difference to image segmentation is that our model also needs to consider the shift of the regions over time. While this does not affect the definitions of the losses, it means that conclusions drawn about the relative merits of loss functions in image segmentation may not be applicable to our problem.

The most common loss for predicting the probability of a binary event with neural networks is the \emph{cross entropy} \citep[CE;][]{Goodfellow2016DeepLearning}, a probability-theoretic measure of the distance between the predicted probability and actual occurrence. The CE between a probability $p \in (0,1)$ and the true event occurrence $y \in \{0,1\}$, with $1$ signifying occurrence of the event and $0$ signifying non-occurrence, can be defined for each pixel as
\begin{eqnarray}
    \mathrm{CE}(p,y) &=& -\log p_\mathrm{t} \\
    p_\mathrm{t} &=& \begin{cases} p, & y = 1 \\ 1-p, & y = 0 .\end{cases}
\end{eqnarray}
The CE weights the two classes equally. With unbalanced datasets, it is common to give a higher weight to the less frequent class, which tends to be more important. In this article, we call this weighted cross entropy (WCE) 
\begin{eqnarray}
    \mathrm{WCE}(p,y) &=& \alpha_\mathrm{t} \mathrm{CE}(p,y) \\
    \alpha_\mathrm{t} &=& \begin{cases} \alpha_1, & y = 1 \\ \alpha_0, & y = 0 .\end{cases}
\end{eqnarray}
where $\alpha_c$ are the weights of the classes $c \in \{0,1\}$. Often, the weights are set proportional to the inverse of the frequency $f_c$ of each class $c$. In this work we set, for both classes,
\begin{equation}
    \alpha_c = (2f_c)^{-1} \label{eq:IFW}
\end{equation}
where $f_c$ is the frequency of that class (with $f_0+f_1=1$). This has the convenient property that the average sample weight is $f_0 \alpha_0 + f_1 \alpha_1 = 1$. The weighting comes at the cost that, unlike CE, WCE is no longer a proper distance measure between the probability distributions.

CE considers all points equally regardless of the difficulty of classifying them. \citet{Lin2017FocalLoss}, arguing that a loss function should focus more on the pixels whose classification is more uncertain, introduced the focal loss (FC) and its weighted version (WFC), which can be defined as
\begin{eqnarray}
    \mathrm{FC}(p,y) &=& (1-p_\mathrm{t})^\gamma \mathrm{CE}(p,y) \\
    \mathrm{WFC}(p,y) &=& (1-p_\mathrm{t})^\gamma \mathrm{WCE}(p,y)
\end{eqnarray}
where $\gamma \geq 0$ is a focusing parameter. With $\gamma=0$, FL reduces to CE, while larger values of $\gamma$ upweight the more uncertain cases, in which $p_\mathrm{t}$ is smaller, relative to the less uncertain ones. In image segmentation tasks, FL has been found to perform better than CE \citep{Chang2018FocalLoss,Doi2018FocalLoss}.

If the performance of the model is judged in terms of one of the metrics presented in Sect.~\ref{sect:models}\ref{sect:evaluation}, it is reasonable to attempt to train the model explicitly to optimize that metric. Since metrics like those given in Eqs.~\ref{eq:CSI}--\ref{eq:PSS} are based on a sharp threshold, they cannot be used directly to train neural networks, which require a differentiable loss function. We can create differentiable analogues of these metrics by considering the definitions of TP, FP, FN and TN in Eqs.~\ref{eq:tp}--\ref{eq:tn} and replacing $\neg$ and $\land$ by differentiable versions
\begin{eqnarray}
    \neg p &\coloneq& 1 - p \\
    p \land q &\coloneq& pq.
\end{eqnarray}
If $p$ and $q$ are discrete logical values represented by $0$ and $1$, these are equivalent to the logical definitions, but they also function in a differentiable manner for continuous $p,q \in [0,1]$. Thus, it is possible to convert CSI, HSS or PSS, or any of the similar skill scores, to a loss function, although a small smoothing factor in the denominator is necessary to prevent division by zero. In this work, we consider the case of CSI loss which has been adopted in image segmentation as the \emph{intersection-over-union loss} \citep{Rahman2016IOU}.

\section{Results and discussion} \label{sect:results}

\subsection{Model selection} \label{sect:selection}

In order to refine the network, we experimentally evaluated the effect of various design choices on its skill. All such evaluations were carried out using the validation dataset; the test set was set aside for the evaluation of the final selected model. Many hyperparameter choices were already examined for a similar network by \citet{Leinonen2021Weather4castBigData}, though this was done in the context of predicting a continuous variable using mean square error loss. Hence, the main focus of model tuning in this work was on adapting the network to probabilistic predictions.

\subsubsection{Loss functions}

The most important difference between the continuous predictions in 
\citet{Leinonen2021Weather4castBigData} and the binary categorical predictions in this work is the choice of loss function. We evaluated the performance of the model using the various losses introduced in Sect.~\ref{sect:models}\ref{sect:loss}. Each loss was evaluated using two choices of class weighting: equal weighting and inverse frequency weighting (IFW; Eq.~\ref{eq:IFW}) setting the occurrence of the target variable to $f_1=0.0106$. An exception is the CSI loss, which is naturally weighted and accordingly was tested only once. 

\begin{figure}
    \ifdefined\ARXIV
        \centerline{\includegraphics[width=\linewidth]{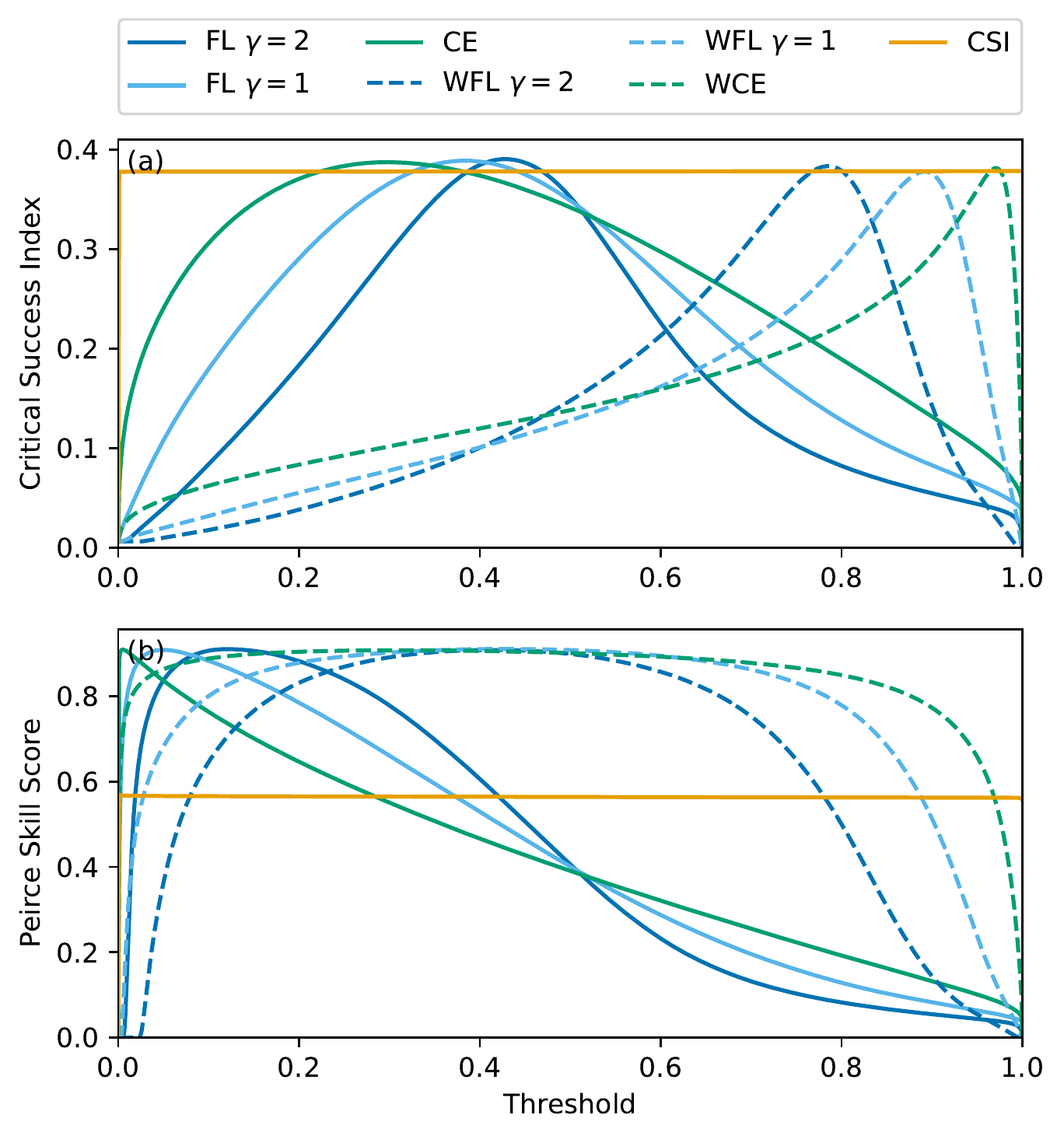}}
    \else
        \centerline{\includegraphics[width=0.7\textwidth]{metrics-loss.pdf}}
    \fi
    \caption{(a) CSI as a function of threshold skill scores for different loss functions. (b) As panel a, but with PSS.}
    \label{fig:metrics-loss}
\end{figure}

In Fig.~\ref{fig:metrics-loss}, we show the CSI and PSS metrics for the different losses as a function of the threshold chosen. The ETS and HSS metrics, while numerically different from CSI, exhibit essentially the same patterns as CSI, hence they are omitted from Fig.~\ref{fig:metrics-loss}. FL with $\gamma=2$ attains the highest of both scores, with CSI of $0.391$ and PSS of $0.910$. The differences in the top performance scores between the loss choices are modest except for the CSI loss, which produces a fairly good CSI but performs poorly with PSS. The choice of loss strongly affects the threshold $T$ where the metrics peak. With both CSI and PSS, the equally weighted models peak at lower $T$ than the IFW models. Increasing $\gamma$, i.e. giving more weight to the uncertain cases, increases the optimal $T$ with the equally weighted losses but decreases it with the IFW losses. The CSI loss tends to produce outputs that are very close to either $0$ or $1$ with few values between, so the choice of $T$ has little effect on the metrics.

\begin{figure}
    \ifdefined\ARXIV
        \centerline{\includegraphics[width=\linewidth]{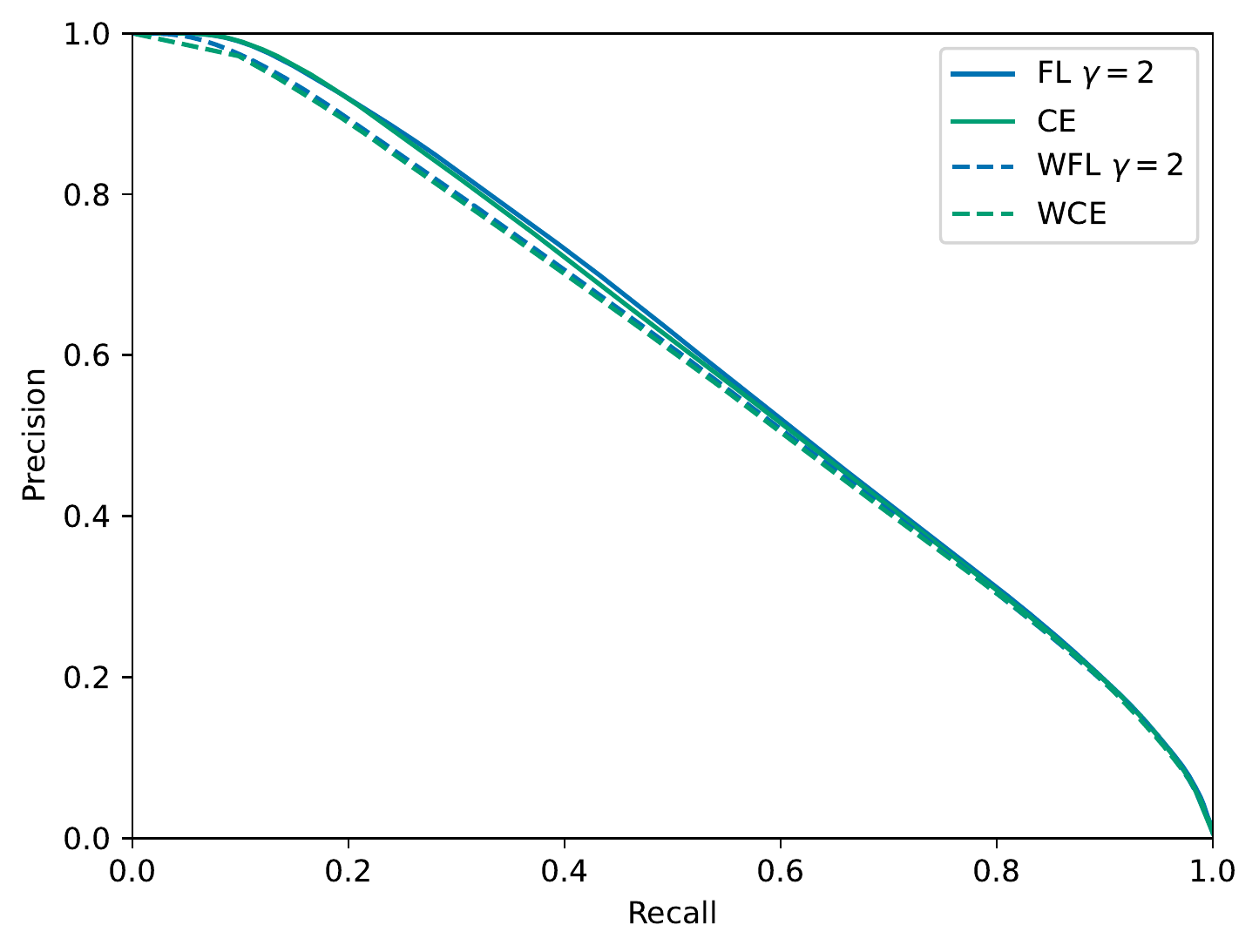}}
    \else
        \centerline{\includegraphics[width=0.7\textwidth]{pr-loss.pdf}}
    \fi    
    \caption{Precision-recall curves for various loss functions.}
    \label{fig:pr-loss}
\end{figure}

The equally weighted losses outperform the IFW losses. We find this result somewhat surprising, as IFW has been established as relatively standard way of helping models learn from unbalanced datasets and is endorsed by, e.g., the original paper on FL \citep{Lin2017FocalLoss}. The result is further supported by Fig.~\ref{fig:pr-loss}, where we show the precision-recall curves for different models. These curves also show the equally weighted losses slightly outperforming the IFW losses. For visual clarity, we have omitted the $\gamma=1$ losses from Fig.~\ref{fig:pr-loss} as their precision-recall curves are indistinguishable from the CE and $\gamma=2$ equivalents. The CSI loss has also been omitted as its tendency to yield values very close to $0$ or $1$ makes it difficult to produce precision-recall curves.

\subsubsection{Calibration} \label{sect:calibration}

In order to yield proper probabilistic forecasts, the output of the model must correctly reflect the probability of the event occurring. Among the loss functions presented in Sect.~\ref{sect:models}\ref{sect:loss}, only CE is strictly a probability-theoretic metric of the distance between distributions. Accordingly, this is the only loss that we can expect to accurately represent the probability of an event occurring in a given pixel. Both the IFW and the additional factor introduced in FL break the probabilistic assumptions.

\begin{figure}
    \ifdefined\ARXIV
        \centerline{\includegraphics[width=\linewidth]{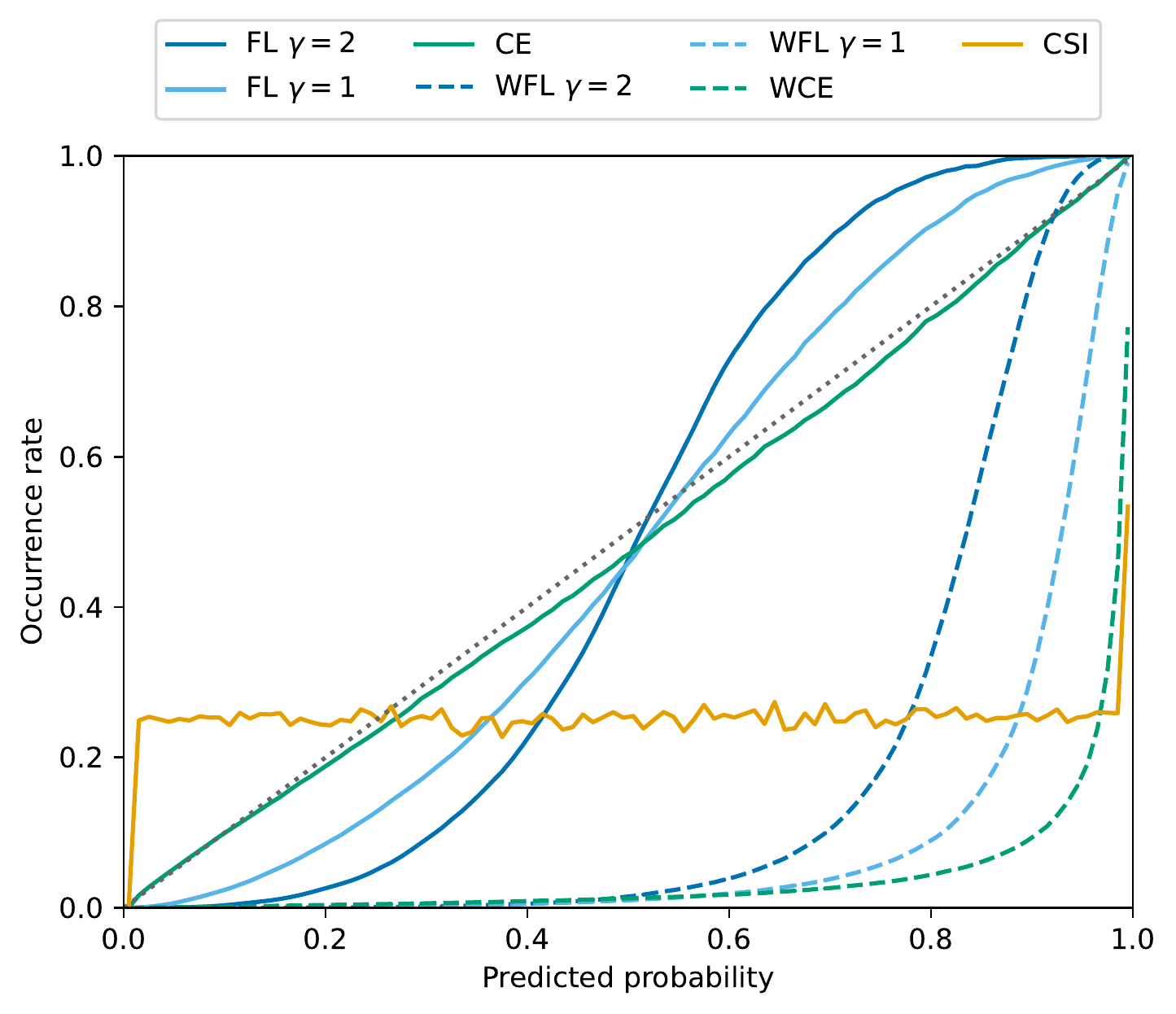}}
    \else
        \centerline{\includegraphics[width=0.7\textwidth]{calibration-loss.pdf}}
    \fi
    \caption{Model calibration. The probability predicted by the model is shown on the horizontal axis and the actual rate of occurrence for that prediction on the vertical axis. The dotted gray line indicates $1:1$ correspondence.}
    \label{fig:calibration-loss}
\end{figure}

Calibration curves, which express the occurrence rate of the target variable as a function of the predicted probability $p$, are shown in Fig.~\ref{fig:calibration-loss} for the different losses. The occurrence rate has been calculated for $100$ bins equally spaced in $p$. As expected, only the CE loss produces a near $1:1$ correspondence between the predicted and observed occurrence. Equally weighted FL results in a roughly sigmoid-shaped curve that crosses the $1:1$ line near $0.5$. The IFW losses result in calibration curves that remain low until rather high $p$, in particular for WCE, and then increase steeply. The output of the CSI loss is very heavily weighted towards values near $0$ and $1$, producing sampling noise in the other bins. 

It is possible to recalibrate models by applying the curves shown in Fig.~\ref{fig:calibration-loss} to the model output. The recalibration does not affect metrics like CSI and PSS or the precision-recall curves because it merely shifts the thresholds. Recalibration is easier if the curve properly covers the full $[0,1]$ range of occurrence rates and if it is not too steep at any point. The WCE has a steep calibration curve near $p=1$, which complicates the calibration, while the CSI loss is essentially impossible to calibrate. The focal losses produce usable calibration curves for both the IFW and equally weighted variants. Meanwhile, the equally weighted CE loss can probably be used without recalibration in most applications.

It is interesting to contrast the results shown here to the conclusions of \citet{Mukhoti2020FocalCalibration}, who argued that FL improves the calibration of the models over CE. We find that this is true for the IFW losses, but the opposite happens with the equally weighted losses where increasing $\gamma$ makes the calibration worse.

\subsubsection{Dropout} \label{dropout}

Dropout and weight regularization are commonly used to prevent overfitting and improve convergence in the training of neural networks. Dropout was omitted by \citet{Leinonen2021Weather4castBigData} as the model architecture was found not to be very prone to overfitting. In this work, we examined this in more detail. We experimented with two alternatives: an FL2 model using dropout with a rate of $0.1$ in the downsampling and upsampling layers as well as weight decay with a rate of $10^{-4}$ in the AdaBelief optimizer, and another without these features. For both alternatives, we trained three instances which were trained identically except for the random initialization of weights.

The CSI, PSS and PR AUC scores, and their standard deviations, are shown in the first two rows of Table~\ref{table:valid-metrics}. They indicate that the models using dropout (DO) and weight decay (WD) perform slightly better than the models without them, although the differences are only on the order of one standard deviation. The standard deviation values also suggest that using these features improves training stability, producing more consistent scores across different training runs. Therefore, we elected to use dropout and weight decay in this study. 
\begin{table*}
\caption{Skill score comparison of different models with the validation set. The numbers indicate the mean score; if a $\pm$ sign is present, the following value is the sample standard deviation. DO stands for dropout and WD for weight decay. The skill scores are determined with different thresholds $T$ optimizing each skill score independently.} \label{table:valid-metrics}
\begin{center}
\begin{tabular}{lccc}
\topline
 & CSI & PSS & PR AUC \\
\midline
FL2 without DO+WD & $0.387 \pm 0.0032$ & $0.908 \pm 0.0006$ & $0.600 \pm 0.0039$ \\
FL2 with DO+WD & $0.389 \pm 0.0013$ & $0.909 \pm 0.0023$ & $0.606 \pm 0.0016$ \\
FL2 ensemble & $0.398$ & $0.913$ & $0.617$ \\
\botline
\end{tabular}
\end{center}
\end{table*}

\subsubsection{Model variance and ensembling}

We selected the model using FL with $\gamma=2$ with dropout and weight decay as our primary model based on its good performance with the metrics and on the ease of recalibrating it. Since three instances of the model had been trained for examining the effect of DO and WD, we also created an ensemble model that outputs the average of of the three models. Such ensembling is often found to result in performance that is better than that of any of the individual models \citep{Ganaie2021Ensemble}. We obtained the same result; the ensemble scores shown on the last row of Table~\ref{table:valid-metrics} demonstrate that the ensemble outperforms the individual models.

\subsection{Evaluation}

Having selected the model based on the results of Sect.~\ref{sect:results}\ref{sect:selection}, we use the recalibrated ensemble of three FL $\gamma=2$ models to evaluate results on the test dataset. 

\subsubsection{Skill metrics}

Various skill scores of our model are shown in Table~\ref{table:metrics} and compared to the Eulerian and Lagrangian persistence models. The Eulerian persistence model assumes that lightning activity including its location remains the same as on the last time step in the past. In the Lagrangian persistence model, the lightning field from the final past time step is additionally motion extrapolated using motion detected from the RZC field using the Lucas--Kanade method \citep{LucasKanade1981Vision} implemented in the Pysteps library \citep{Pulkkinen2019Pysteps}. The skill scores were calculated by selecting the threshold $T$ such that it gives the optimal CSI in the validation dataset, then evaluating the scores using the test dataset. The optimal $T$ for the test dataset would have been only slightly different, $0.421$ instead of $0.426$. The skill scores other than CSI in Table~\ref{table:metrics} were also computed using this $T$ even though other choices of $T$ may be more optimal if one wishes to optimize other skill scores instead. This is in contrast to Table~\ref{table:valid-metrics}, where each score was computed using the $T$ optimal for that score. Consequently, the PSS is worse in Table~\ref{table:metrics} than in Table~\ref{table:valid-metrics}, but since $T$ was selected differently, this does not indicate that the model performs worse with the test dataset than with the validation set. Indeed, the better CSI score for the test set ($0.453$) compared to the validation set ($0.398$) indicates that the test set is somewhat less challenging for the model. Regardless, our model clearly outperforms the persistence models for all metrics. The AUC scores cannot be computed for the persistence models as they are not probabilistic.
\begin{table*}
\caption{Skill scores of the model on the test dataset. The scores for the Lagrangian and Eulerian persistence models are shown for comparison. The threshold $T$ was selected to give the optimal CSI for the validation set.} \label{table:metrics}
\begin{center}
\begin{tabular}{lccccccccc}
\topline
 & $T$ & POD & FAR & CSI & ETS & HSS & PSS & ROC AUC & PR AUC \\
\midline
Model & $0.426$ & $0.610$ & $0.362$ & $0.453$ & $0.449$ & $0.620$ & $0.607$ & $0.989$ & $0.688$ \\
Lagrangian & --- & $0.473$ & $0.509$ & $0.317$ & $0.313$ & $0.476$ & $0.468$ & --- & --- \\
Eulerian & --- & $0.439$ & $0.581$ & $0.273$ & $0.268$ & $0.422$ & $0.432$ & --- & --- \\
\botline
\end{tabular}
\end{center}
\end{table*}

Comparisons of the skill metrics to the results of earlier studies, whether ML-based or traditional, are difficult to perform because their definitions for lightning occurrence differ from our ``within $\mathrm{8}\ \mathrm{km}$ in the last 10 min'' definition. Some studies also calculate the metrics using each thunderstorm cell as a data point, while we use a more demanding point-wise calculation that is likely to produce lower scores. Hence, we avoid a direct comparison here.

\subsubsection{Effect of lead time on skill}

\begin{figure}
    \ifdefined\ARXIV
        \centerline{\includegraphics[width=\linewidth]{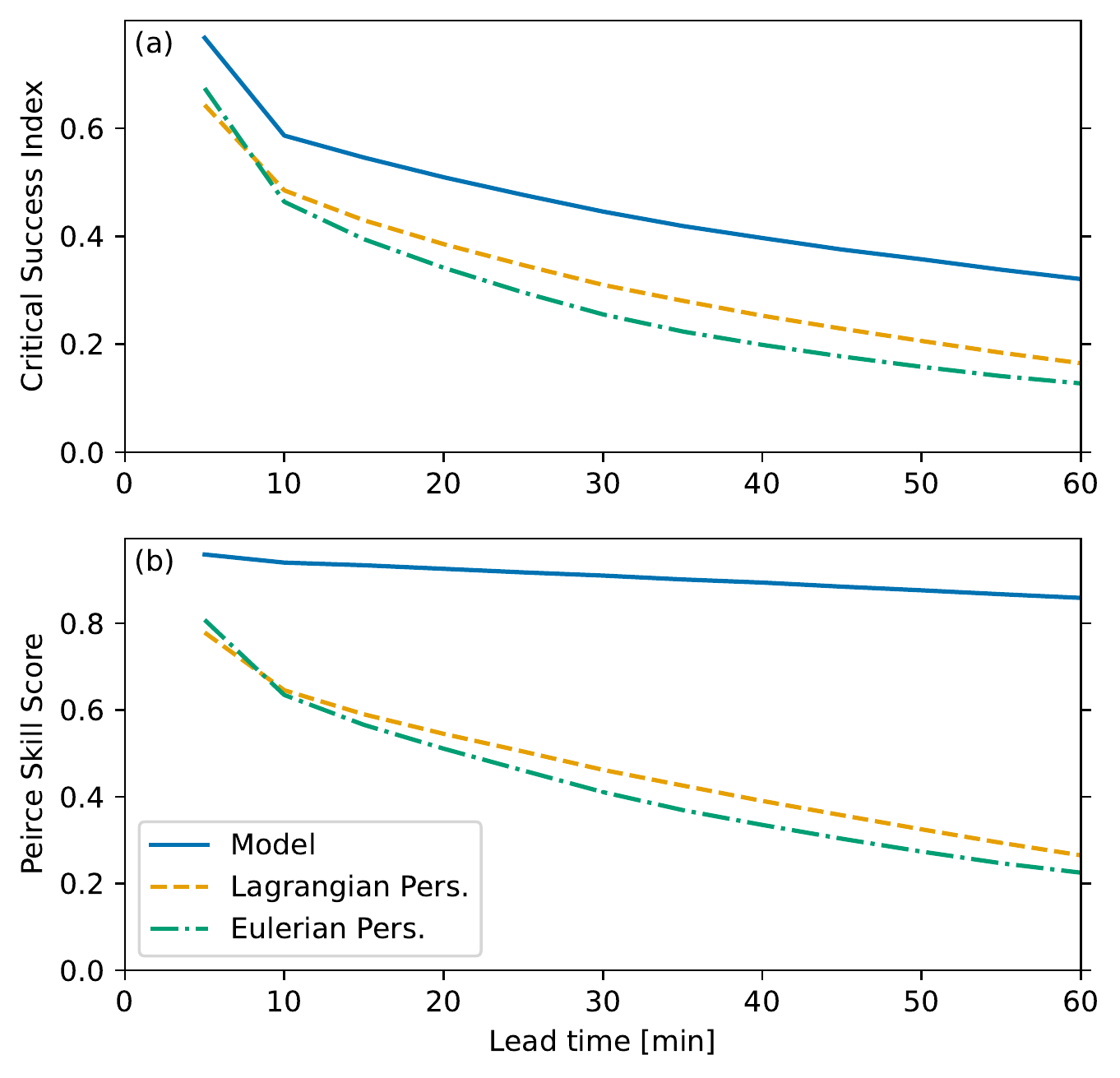}}
    \else
        \centerline{\includegraphics[width=0.7\textwidth]{metrics-leadtime.pdf}}
    \fi
    \caption{(a) CSS and (b) PSS of the model (blue) at lead times from 5 to 60 min, compared to the Lagrangian (orange) and Eulerian (green) persistence assumptions.}
    \label{fig:metrics-leadtime}
\end{figure}

The skill of a forecast model can be expected to degrade with increasing lead time. In Fig.~\ref{fig:metrics-leadtime}, we show the optimal CSI and PSS for the model as a function of lead time between $5\ \mathrm{min}$ and $60\ \mathrm{min}$. These are compared to the equivalent results of the Eulerian and Lagrangian persistence models. The performance at the first time step with $5\ \mathrm{min}$ lead time is high as the target variable is lightning occurrence within the last $10\ \mathrm{min}$, meaning that some of the correct answers on the first time step can be inferred directly from the input. There is a rather sharp drop in CSI from 5 min to 10 min, followed by a more gradual decline. The relative advantage of our deep-learning model over the persistence model grows with increasing lead time. With PSS, our model has rather high scores even at long lead times because PSS weights detections much more than false negatives.

\subsubsection{Example cases}

While it is not possible to cover the wide variety of possible cases within the constraints of this article, we chose three examples for discussion that demonstrate the ability of the model to predict the movement, growth and decay of convective systems. These have all been evaluated with the calibrated model such that the predicted probabilities accurately reflect the probability of occurrence.

\begin{figure*}
    \centerline{\includegraphics[width=\textwidth]{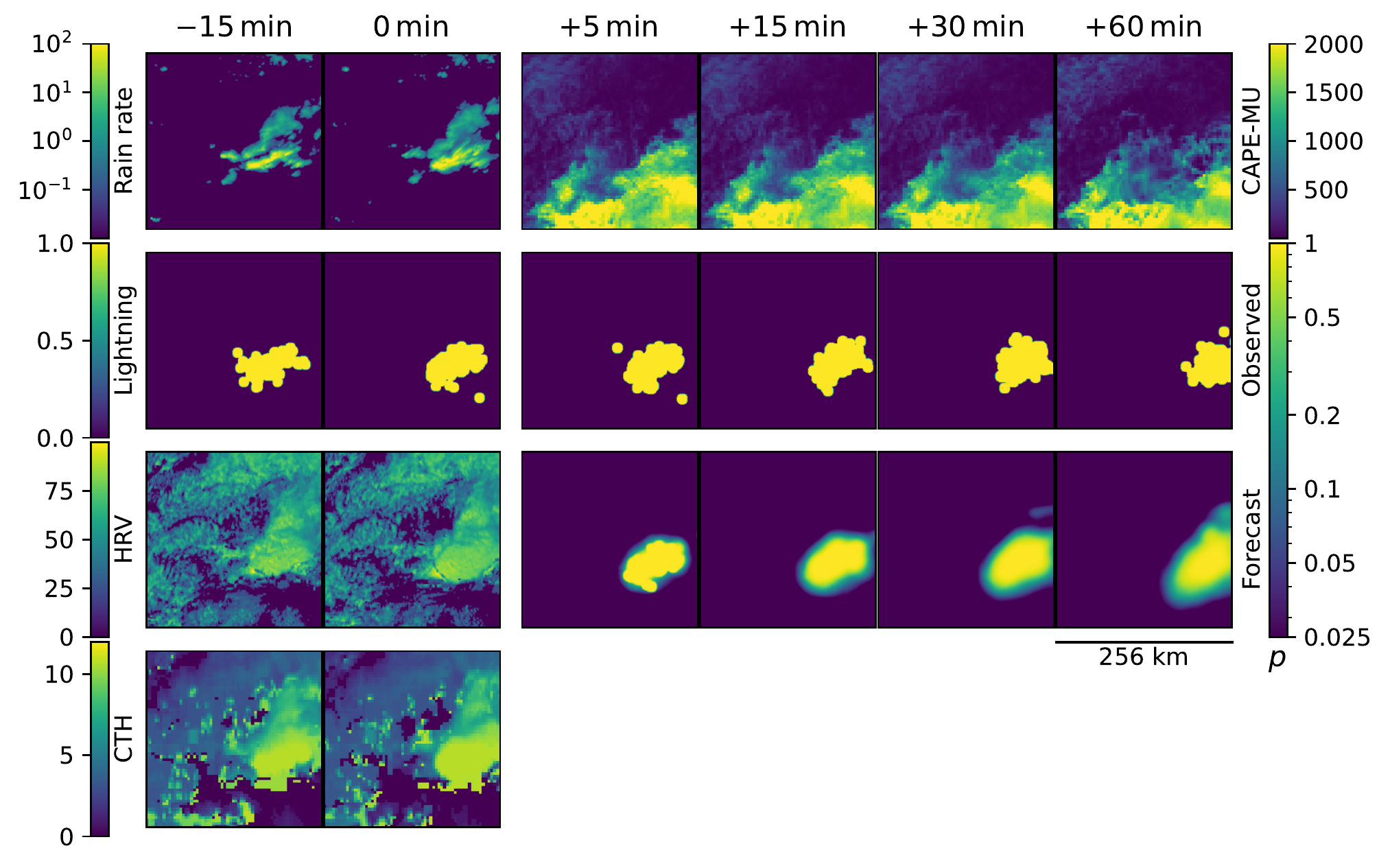}}
    \caption{An example of predictions with our model in a case with a moving convective system on July 11, 2020 at 12:00 UTC. The two columns of images on the left show four input variables: rain rate, lightning occurrence as defined for the target variable, the HRV satellite image and cloud top height (CTH). The four columns on the right show the NWP-predicted CAPE, observed lightning occurrence and the lightning probability predicted by our model at lead times indicated on top of each column.}
    \label{fig:model-example-0}
\end{figure*}

The first example, shown in Fig.~\ref{fig:model-example-0}, shows a relatively fast-moving system that is actively producing lightning. Comparing the observed and forecast lightning activity demonstrates that the model has correctly inferred the speed and direction of the motion of the system. It also correctly predicts that lightning activity in the system will continue at a similar intensity, giving high confidence that lightning will occur on the center right of the image even at the last time step of the prediction at $t=+{60}\ \mathrm{min}$.

\begin{figure*}
    \centerline{\includegraphics[width=\textwidth]{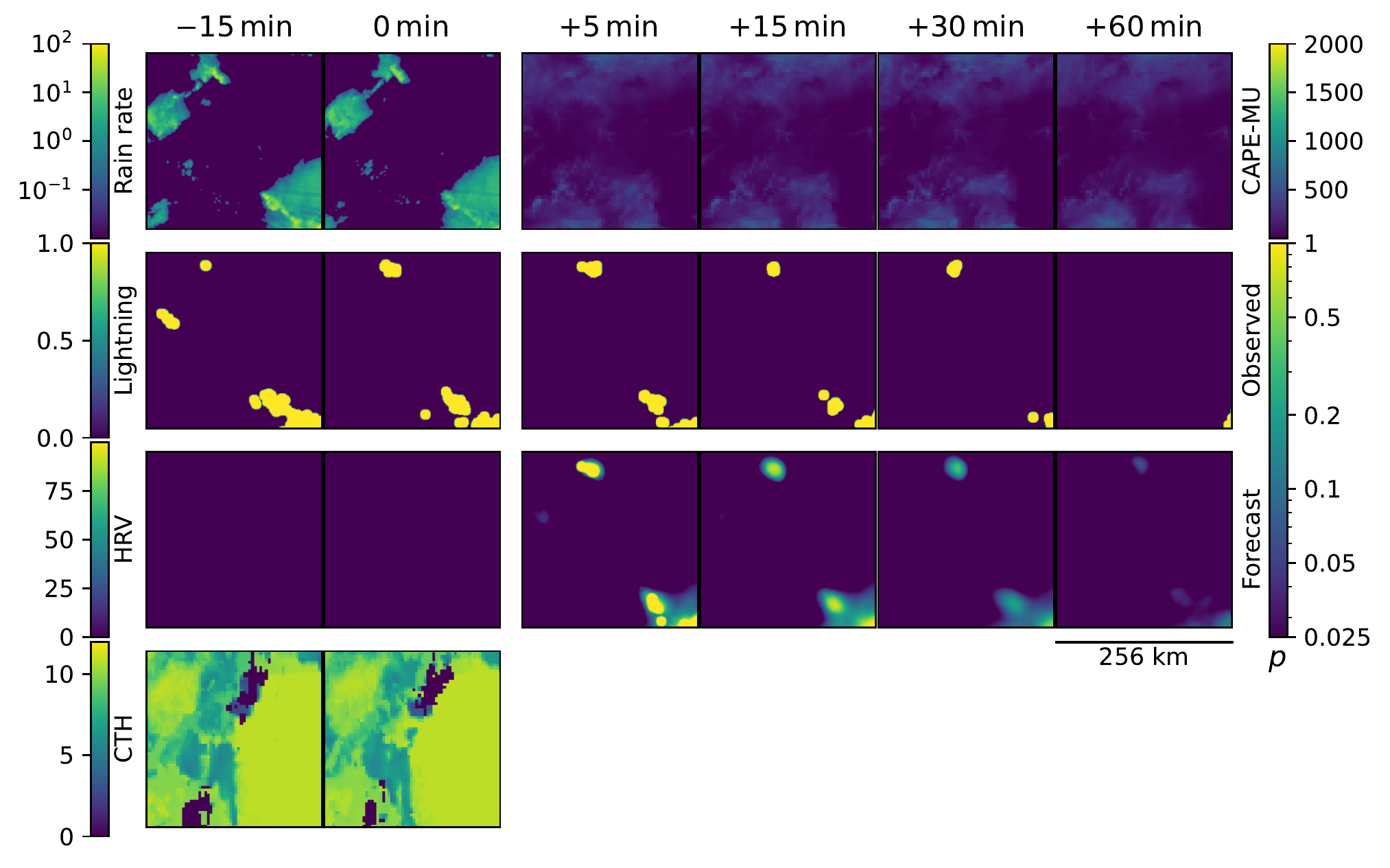}}
    \caption{As Fig~\ref{fig:model-example-0}, but with decaying thunderstorms on August 2, 2020 at 01:30 UTC. The HRV data are missing as the case occurs at night.}
    \label{fig:model-example-1}
\end{figure*}

In the second example in Fig.~\ref{fig:model-example-1}, the lightning activity decreases from $t \leq 0$ to $t > 0$. The model is able to recognize this from the input sequence and correctly predicts decreasing probabilities in all regions of lightning activity. It forecasts very little lightning activity for the last time step, and this is indeed the case in the observation. 

\begin{figure*}
    \centerline{\includegraphics[width=\textwidth]{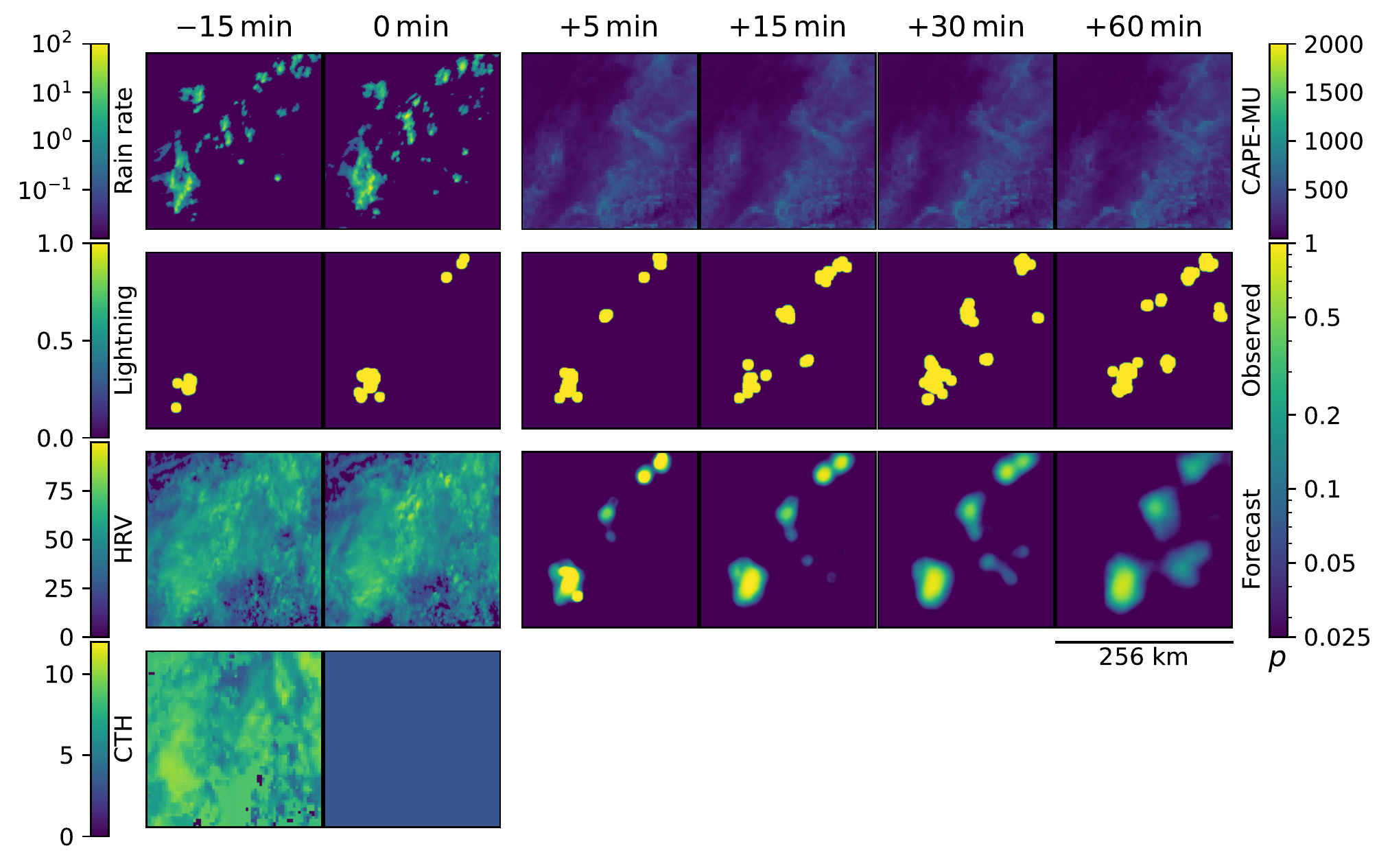}}
    \caption{As Fig~\ref{fig:model-example-0}, but with growing thunderstorms on August 2, 2020 at 10:15 UTC. The data are missing for the $t=0$ frame of the CTH.}
    \label{fig:model-example-2}
\end{figure*}

In contrast to the previous example, Fig.~\ref{fig:model-example-2} shows initiating and intensifying cells. Three slightly different cases can be identified. First, the cell on the bottom left is already active at $t=-{15}\ \mathrm{min}$ and the model predicts with high confidence that it will also remain active. Second, on the top right lightning activity has just initiated, being present at $t=0$ but not at $t=-{15}\ \mathrm{min}$. The model infers that activity in this area will continue. Finally, near the center there is an area where no lightning has been detected in the input data. Nevertheless, the model detects a growing cell in this area, presumably using other input variables, and correctly predicts that lightning activity will begin there. There is a region on the center right of this example where lightning occurs in the observation while the predicted probability does not exceed the $p=0.025$ threshold for visualization. However, the assigned probability is still nonzero in this area, ranging between approximately $0.01$ and $0.02$ at $t=+{60}\ \mathrm{min}$. In contrast, in the top left corner that is farthest from lightning activity, the predicted probability is approximately $3 \times 10^{-5}$, indicating much higher confidence in the absence of lightning.

As manually selected cases, the examples shown in this section are not statistically representative of the dataset; to provide a representative sample, we have included equivalent figures of $32$ random cases from the test dataset in the accompanying data archive \citep{Leinonen2022Weights}. These cases, or indeed the entire test dataset, cannot fully cover the variety of cases that our model might encounter, especially if it is used in conditions that are considerably outside the training distribution. However, they demonstrate that the network works well in a wide variety of cases and does not easily suffer from artifacts or glitches creating spurious lightning predictions.

\section{Conclusions} \label{sect:conclusions}

The ability of deep neural networks to learn spatial relationships using convolutional layers and temporal relationships using recurrent layers makes recurrent-convolutional networks, which utilize both of these, a natural fit for forecasting the spatiotemporal evolution of atmospheric fields. The network introduced in this study utilizes this architecture to estimate the probability of lightning occurrence at each grid point and time step in the 60 min following the reference time. It uses inputs from ground-based radar, satellite observations, lightning detection, numerical weather prediction and a digital elevation model. The output probability can be utilized flexibly by the end users to issue warnings at specific thresholds, balancing their tolerance to non-detections and false alarms.

The results indicate that our selected model is able to infer the stage of the lifecycle of convection from the input data. It can predict the motion, growth and decay of lightning-producing thunderstorm cells, adding to its ability to accurately predict the occurrence of lightning. The ability to directly predict the movement using only convolutional and recurrent layers also means that object detection and tracking are not necessarily needed to nowcast thunderstorm hazards.

Ideally, a probabilistic nowcasting model for the occurrence of rare events should separate the occurrences from non-occurrences as efficiently as possible, and accurately represent the probability of the event occurring. Whether the latter is the case depends on the choice of loss function. Among the losses we examined, only the cross entropy is strictly probabilistic, while others, such as the focal loss or losses utilizing unequal class weighting, break the assumptions of the probabilistic loss. The other losses require recalibration of the output in order to be interpreted as a probability. In this work, we adopted the focal loss with focusing parameter $\gamma=2$, which does require recalibration but whose calibration curve is well-behaved enough that this can be done easily. While this loss achieved the best metrics, cross entropy performed only slightly worse and is naturally calibrated, making it a good alternative in cases where recalibration is not practical or desirable. Contrary to established practice, we found that equal weighting of the lightning occurrence and non-occurrence classes produced better results than inverse frequency weighting even though the dataset is severely unbalanced.

Given that our network architecture is not specific to lightning, and that we exploit a multi-source dataset in this study that can give us information about many different hazards, it is expected that our approach can be easily adapted to other input and target variables. For instance, the upcoming Meteosat Third Generation satellites will provide higher-resolution geostationary observations for Europe, potentially helping CNNs extract more information. Furthermore, the importance of different data sources, previously examined by \citet{Zhou2020LightningDL} and \citet{Leinonen2021DataSources}, is yet to be quantified in this context, but is necessary in order to understand the expected performance of the network in, for example, regions where ground-based radar observations are not available. We intend to investigate these topics in detail in a follow-up study. Further input variables could also be added in future versions, such as a distinction between cloud-to-ground and cloud-to-cloud lightning, polarimetric radar variables, observed or simulated hydrometeor densities that can act as indicators of lightning activity \citep{Besic2016Hydrometeor,Figueras2019LightningRadar}, and a more detailed description of the planetary boundary layer.

Finally, we found it difficult to compare our results to those of other studies due to differences in the datasets and the lightning occurrence definitions. To remedy this, we recommend that the community adopt standardized definitions and benchmark datasets in order to enable fair comparisons between different approaches.

\clearpage

\acknowledgments
We thank Simone Balmelli for his assistance with the lightning data. The work of JL was supported by the fellowship ``Seamless Artificially Intelligent Thunderstorm Nowcasts'' from the European Organisation for the Exploitation of Meteorological Satellites (EUMETSAT). The hosting institution of this fellowship is MeteoSwiss in Switzerland.

%
%
\datastatement
The preprocessed training, validation and testing datasets created for this study, as well as the trained models and precomputed results are available for noncommercial use under the CC BY-NC-SA 4.0 license at \url{https://doi.org/10.5281/zenodo.6802292} \citep{Leinonen2022Weights}. The ML and analysis code used in this study can be found at \url{https://github.com/MeteoSwiss/c4dl-lightningdl}. The original data from EUCLID lightning network are proprietary and cannot be made available in raw form. The original data from the Swiss radar network and the COSMO NWP model can be made available for research purposes on request. The MSG SEVIRI Rapid Scan radiances are available to EUMETSAT members and participating organizations at the EUMETSAT Data Store (\url{https://data.eumetsat.int/}). The NWCSAF products can be created from these data using the publicly available NWCSAF software available at \url{https://www.nwcsaf.org/}. The ASTER DEM can be obtained from \url{https://doi.org/10.5067/ASTER/ASTGTM.003} \citep{ASTERGDEMV3Data}.

\appendix[A]
\appendixtitle{Input variables}
Lists of input variables and details on their preprocessing are provided in Tables~\ref{table:preprocessing-1} and~\ref{table:preprocessing-2}.
\clearpage
\begin{table*}
    \caption{Summary of the data variables and their preprocessing. The abbreviations are explained in Sect.~\ref{sect:data}.} \label{table:preprocessing-1}
    \begin{center}
    \begin{tabular}{l|c|c}
        \textbf{Variable} & \textbf{Fill} & \textbf{Transform} \\
        \hline
        \multicolumn{3}{l}{\textbf{Lightning}} \\
        Density & ${10^{-4}}\ \mathrm{km^{-2}}$ & $(\log_{10}(x) + 0.593)/0.640$ \\
        Current & ${10^{-8}}\ \mathrm{kA\,km^{-2}}$ & $(\log_{10}(x) - 0.0718)/0.731$ \\
        8 km / 10 min occurrence & --- & --- \\
        
        \multicolumn{3}{l}{} \\
        \multicolumn{3}{l}{\textbf{Radar}} \\
        RZC & ${0.01}\ \mathrm{mm\,h^{-1}}$ & $(\log_{10}(x) + 0.051)/0.528$ \\
        CZC & ${-5}\ \mathrm{dBZ}$ & $(x-{21.3}\ \mathrm{dBZ})/{8.71}\ \mathrm{dBZ}$ \\
        LZC & ${0.5}\ \mathrm{g\,m^{-3}}$ &  $(\log_{10}(x) + 0.274)/0.135$ \\
        EZC-20, EZC-45, HZC & $0$ & $x / {1.97}\ \mathrm{km}$ \\
        
        \multicolumn{3}{l}{} \\
        \multicolumn{3}{l}{\textbf{Satellite}} \\
        \makecell[l]{Solar:\\HRV, VIS006,\\VIS008, IR-016} & --- & $x/{100}\ \mathrm{K}$ \\
        Solar/Thermal: IR-039 & --- & $(x - {274}\ \mathrm{K})/{17.5}\ \mathrm{K}$ \\
        \makecell[l]{Thermal:\\WV-063, WV-073,\\IR-087, IR-097,\\IR-108, IR-120, IR-134} & --- & $(x - {250}\ \mathrm{K})/{10}\ \mathrm{K}$ \\
        Cloud top temperature & ${330}\ \mathrm{K}$ & $(x - {260}\ \mathrm{K})/{19.1}\ \mathrm{K}$ \\
        Cloud top height & ${-1000}\ \mathrm{m}$ & $(x - {5260}\ \mathrm{m})/{2810}\ \mathrm{m}$ \\
        Cloud top phase & --- & ---\\
        Cloud optical thickness & $0.1$ & $(\log_{10}(x)-0.94)/0.588$\\
    \end{tabular}
    \end{center}
\end{table*}
\clearpage

\clearpage
\begin{table}
    \caption{Continued from Table~\ref{table:preprocessing-1}.} \label{table:preprocessing-2}
    \begin{center}
    \begin{tabular}{l|c|c}        
        \textbf{Variable} & \textbf{Fill} & \textbf{Transform} \\
        \hline
        \multicolumn{3}{l}{\textbf{NWP}} \\
        CAPE & --- & $x/200\ \mathrm{J\,kg^{-1}} $ \\
        CIN & --- & $x/21\ \mathrm{J\,kg^{-1}}$ \\
        HZEROCL & $0$ & $x/3300\ \mathrm{m}$ \\
        LCL & --- & $x/1000\ \mathrm{m}$ \\
        MCONV & --- & $x/3.8 \times 10^{-6}\ \mathrm{s^{-1}}$ \\
        OMEGA & --- & $x/4.2\ \mathrm{Pa\,s^{-1}}$ \\
        SLI & --- & $x/3.5\ \mathrm{K}$ \\
        Soil type & --- & --- \\
        T-2M, T-SO & --- & $(x - 290\ \mathrm{K})/7.2\ \mathrm{K}$ \\
        
        \multicolumn{3}{l}{} \\
        \multicolumn{3}{l}{\textbf{DEM}} \\
        Altitude & --- & $x / 820\ \mathrm{m}$ \\
        \makecell[l]{East--west derivative,\\North--south derivative} & --- & $x / 200\ \mathrm{m\,km}^{-1}$\\
        
        \multicolumn{3}{l}{} \\
        \multicolumn{3}{l}{\textbf{Auxiliary}} \\
        Solar zenith angle & --- & --- 
    \end{tabular}
    \end{center}
\end{table}
\clearpage

\bibliographystyle{ametsocV6}
\bibliography{journalabrv,lightningrdl}

\end{document}